\PassOptionsToPackage{pdfpagelabels=false}{hyperref}
\documentclass[fleqn, usenatbib]{mnras}
\usepackage{newtxtext,newtxmath}
\usepackage[T1]{fontenc}
\usepackage{ae,aecompl}
\usepackage{graphicx}	
\usepackage{amsmath}
\usepackage{multicol}        
\usepackage{bm}	
\usepackage{pdflscape}
\usepackage[usenames]{color}
\usepackage{soul}
\usepackage{wrapfig}
\usepackage[utf8]{inputenc}
\usepackage[section]{placeins}
\usepackage{float}
\usepackage{pgfplots}
\usepackage{natbib}
\usepackage[nottoc]{tocbibind}
\usepackage{color,soul}
\usepackage{enumerate}
\usepackage{caption}
\usepackage{subcaption}
\usepackage{verbatim}
\usepackage{graphicx}
\usepackage{subcaption}
\pgfplotsset{compat=newest}

\title{The Force Explosion Condition is Consistent with Spherically Symmetric CCSN Explosions}
\author[M. Gogilashvili \& J.W. Murphy \& E. P. O’Connor]{Mariam Gogilashvili,$^{1,}$$^{3}$ \thanks{Email: mg18u@my.fsu.edu} \, 
Jeremiah W. Murphy,$^{1}$ \thanks{Email: jwmurphy@fsu.edu} \, 
Evan P. O’Connor, $^{2}$  \thanks{Email:  evan.oconnor@astro.su.se}
\\
$^{1}$Department of Physics, Florida State University, 77 Chieftan Way, Tallahassee, FL 32306, USA; \\
$^{2}$The Oskar Klein Centre, Department of Astronomy, Stockholm University, AlbaNova, SE-106 91 Stockholm, Sweden; \\
$^{3}$ Los Alamos National Laboratory, 87545-1362, Los Alamos, NM, USA;}

\begin{document}
\label{firstpage}
\pagerange{\pageref{firstpage}--\pageref{lastpage}}
\maketitle

\begin{abstract}
    One of the major challenges in Core-collapse Supernova (CCSN) theory is to predict which stars explode and which collapse to black holes. \citet{Gogilashvili2022} derived an analytic force explosion condition (FEC) and showed that the FEC is consistent with CCSN simulations that use the light-bulb approximation for neutrino heating and cooling.  In this follow-up manuscript, we show that the FEC is consistent with the explosion condition when using actual neutrino transport in \texttt{GR1D} simulations \citep{OCONNOR2015}.  Since most 1D simulations do not explode, to facilitate this test, we enhance the heating efficiency within the gain region.  To compare the analytic FEC and radiation-hydrodynamic simulations, this manuscript also presents a practical translation of the physical parameters.  For example: we replace the neutrino power deposited in the gain region, $L_\nu\tau_g$, with the net neutrino heating in the gain region; rather than assuming that $\dot{M}$ is the same everywhere, we calculate $\dot{M}$ within the gain region; and we use the neutrino opacity at the gain radius.  With small, yet practical modifications, we show that the FEC predicts the explosion conditions in spherically symmetric CCSN simulations that use neutrino transport.
\end{abstract}


\begingroup
\let\clearpage\relax
\endgroup
\newpage

\section{Introduction}
Understanding how massive stars die has been one of the major challenges in stellar evolution theory for decades \citep{decadal20}. At the end of their lives, nearly all massive stars undergo core collapse. The collapsing core bounces at the nuclear densities and launches a shock wave. If the blast wave overwhelms the collapsing star, then the star explodes as a core-collapse supernova (CCSN) explosion \citep{LI2011,HORIUCHI2011}. Otherwise, the star fails to explode and collapses to a black hole (BH) \citep{FISCHER2009, OCONNOR2011}. 
 Knowing how and which stars explode has important ramifications for a wide range of astrophysical phenomena. Predicting which stars explode, will significantly impact our understanding of nucleosynthesis \citep{WOOSLEY2002,DIEHL2021}, neutron star formation \citep{BAADE1934A,BURROWS1986}, and black hole formation \citep{FISCHER2009, OCONNOR2011}. Moreover, predicting the neutron star and black hole distributions is a first step in predicting the prevalence of compact remnant mergers and their gravitational wave emission. 
Therefore, the major goal of CCSN theory is to predict which stars explode and which fizzle. This requires developing a theory for the conditions of successful core-collapse supernova explosions.

After bounce, the blast wave almost always stalls into an accretion shock.  The primary challenge in CCSN theory is to understand how the stalled shock revives. The pioneering work of \citet{COLGATE1966} suggested that neutrinos can transport energy from the neutron star to the outer layers. Later, \citet{BETHE1985} introduced the "delayed neutrino mechanism" according to which
after a post-bounce delay of hundreds of milliseconds, electron-type neutrinos and anti-neutrinos transport thermal energy from the PNS, heat the material behind the shock which ultimately may lead to shock revival. The delayed neutrino mechanism fails in most spherical one-dimensional (1D) simulations. While the lowest mass progenitors explode \citep{LIEBENDORFER2001A,LIEBENDORFER2001b,LIEBENDORFER2005,
RAMPP2002,BURAS2003,BURAS2006,THOMPSON2003,KITAURA2006,MULLER2017,
RADICE2017}, most of the models do not explode in 1D without boosting the efficiency of $\nu_e$ and $\overline{\nu}_e$ heating. While spherical simulations do not explode, multi-dimensional simulations explode successfully \citep{LENTZ2015,BRUENN2016,MULLER2015,MULLER2019,VARTANYAN2018,VARTANYAN2021}. It is now clear that multi-dimensional effects such as convection and turbulent dissipation play significant role in aiding explosions \citep{MURPHY2008,MABANTA2018}. Hence, multi-dimensional simulations are crucial in understanding the CCSN mechanism. However, multi-dimensional simulations are computationally very expensive, making it difficult to perform systematic studies. 

In addition to 3D simulations, understanding the explosion conditions through analytic means can provide a deeper understanding of the explosion physics.  There has been a handful of attempts to derive an analytic explosion condition. One of the first attempts was the semi-analytic critical curve of \citet{BURROWS1993}. A key assumption of this work is that the stalled shock may be treated as a steady-state problem.  For example, the mass accretion rate, $\dot{M}$ onto the core changes slowly so that the structure can adjust from one steady-state solution to another. The steady-state assumption turns the CCSN problem into a boundary value problem, and they  solved the ODEs with the inner boundary being at the Neutron Star (NS) and outer boundary at the shock. They found that there is a critical curve in two dimensional parameter space of $L_\nu$ - $\dot{M}$. Below the critical curve, steady-state stalled shock solutions exist while above this curve, there are no steady-state solutions. They suggested that the solutions above the critical curve are explosive. \citet{MURPHY2008} showed that the critical curve is consistent with spherical explosions. Moreover, they found that the critical neutrino luminosity is $\sim 30\%$ less in multi-dimensional simulations. Later, \citet{MABANTA2018} showed that convection and turbulent dissipation are responsible for this $\sim 30\%$ reduction in multi-dimensional simulations. 

Since the introduction of the critical neutrino luminosity condition, there have been many attempts to either understand the origin of this condition or to add multi-dimensional effects.  For example, \citet{SUMMA2016} included the effects of turbulent pressure in the gain region, and derived a generalized condition for the critical neutrino luminosity for axisymmetric explosions.   \citet{YAMASAKI2005},~\citet{JANKA2012},~\citet{JANKA2016},~\citet{SUMMA2018} explored the impact of the effects of rotation on the critical neutrino luminosity. 

In an attempt to derive the critical curve of \citet{BURROWS1993}, \citet{KESHET2012} approximately modeled the boundary value problem.  They showed that the critical neutrino luminosity is related to conditions on the neutrino optical depth, $\tau$.  Both $\tau$ and the density at the surface of the NS, $\rho_\nu$, have a maximum for a specific shock radius. Most importantly, they found that this maximum is a monotonic function of the neutrino luminosity and above a critical neutrino luminosity, there are no stalled shock solutions. However, they approximated the neutrino luminosity by assuming black-body emission at the neutrino sphere. While approximating neutrino luminosity in such a way makes calculations easier, it also excludes the need for using the energy conservation equation. Therefore, this critical condition does not explicitly include the neutrino power deposited in the gain region, which is an important parameter in the explodability. 

In another attempt to understand the physics behind the neutrino-luminosity critical condition, \citet{PEJTA2012} introduced the ante-sonic condition. They explored isothermal accretion and showed that there is a maximum sound speed above which steady-state solutions do not exist.  Suspecting similar physics would apply to adiabatic flows, they then used numerical solutions to show that there is roughly an ante-sonic condition for adiabatic accretion as well. The ante-sonic critical condition for the adiabatic flows is $c_s^2/\varv_{\rm esc}^2 \simeq 0.19 $, where $c_s$ is the adiabatic sound speed and $\varv_{\rm esc}$ is the escape velocity at the shock.  \citet{RIVES2018} analytically derived an ante-sonic condition and \citet{RAIVES2021} extended the ante-sonic condition for the flows with rotation and turbulence. 

Others have suggested more qualitative explosion conditions. \citet{THOMPSON2000} suggested a somewhat intuitive time-scale explosion condition, $\tau_{\rm adv}/\tau_{\rm heat}>1$, which compares the advection timescale $\tau_{\rm adv}$ to the heating timescale $\tau_{\rm heat}$. The explosions are successful if the heating timescale within the gain region is shorter compared to the advection timescale through the gain region. \citet{OCONNOR2011} a proposed new parameter, a compactness parameter for the diagnostic of explodability. They were forcing explosions by adjusting neutrino heating and found that there is a correlation between the explodability of progenitors and compactness of Fe core before collapse, $\xi_M=\frac{M/M_\odot}{R(M_{\rm bary}=M)/1000km}\bigg |_{t=t_{\rm bounce}}$. These explosion conditions are good  order-of-magnitude estimates for explodability.  

\citet{ERTL2016} proposed a somewhat more accurate two-parameter criteria that separates successful explosions from failed explosions. They identified these two parameters to be the normalized enclosed mass for a dimensionless entropy per nucleon of $s=4$ (approximately corresponds to an entropy jump at the Si/O interface), $M_4$, and the normalized mass derivative $\mu_4=(dm/M_\odot)(dr/1000km)|_{s=4}$. They showed that $M_4$ and $\mu_4 M_4$ are indirectly related to the mass accretion rate $\dot{M}$ and neutrino luminosity $L_\nu$.

More recently, \citet{WANG2022} proposed that the two main conditions that for explosion are 1.) a strong density discontinuity at the Si/O interface and/or 2.) a steep density profile. They tested hundreds of 2D simulations and found that $\sim 90\%$ of the time, the maximum ram pressure jump that occurs at the time of Si/O interface determines explodability well.  However, the drop in density profile corresponds to large changes in $\dot{M}$ which can suggest that this explosion condition is closely related to critical neutrino luminosity.   This might explain why this condition only works 90\% of the time.

While the previous studies have provided some insight, either they were not analytic and/or they did not compare their predictions with CCSN simulations.

Over the years, many have proposed different explosion conditions, some qualitative, some numerical. However, the critical curve of \cite{BURROWS1993} has remained one of the most reliably accurate and useful conditions for explodability. Motivated by the critical neutrino luminosity condition, \citet{MURPHY2017} used the same semi-analytic technique and proposed a force explosion condition (FEC). They found that the integral of the momentum equation, $\Psi$ plays an important role in the existence of steady-state solutions. They showed that the parameter $\Psi$ determines the balance (imbalance) of forces and is directly correlated to the shock velocity $\varv_s/\varv_{\rm acc}\approx -1+\sqrt{1+\Psi}$, where $\varv_{\rm acc}$ is the velocity of accreting material onto the shock. They found that the condition $\Psi>0$ corresponds to explosive solutions ($\varv_s>0)$. Compared to the original critical condition of \citet{BURROWS1993}, \citet{MURPHY2017} extended the physical parameter space to $L_\nu$, ~$\dot{M}$,~$M_{\rm NS}$, ~$R_{\rm NS}$ and showed that there are more parameters besides neutrino luminosity and mass accretion rate that determine the explosion condition. \citet{MURPHY2017} checked that the integral condition is consistent with the critical neutrino luminosity condition of \citet{BURROWS1993} and also is consistent with simple hydrodynamic simulations. 

Later, \citet{Gogilashvili2022} analytically derived the force explosion condition (FEC) of \citet{MURPHY2017} for spherical explosions. They started from the fundamental equations of hydrodynamics and found that the analytic Force Explosion Condition (FEC) has a simple form $\tilde{L}_\nu\tau_g - a \tilde{\kappa} > b$ and depends upon two dimensionless parameters only. The first is $\tilde{L}_\nu \tau_g = L_{\nu} \tau_g R_{\rm NS}/ ( G \dot{M} M_{\rm NS})$ and compares the neutrino power deposited in the gain region with the accretion power. The second is $\tilde{\kappa} = \kappa \dot{M} / \sqrt{G M_{\rm NS} R_{\rm NS}}$ and parameterizes the neutrino optical depth in the accreted matter near the neutron-star surface.  $a$ and $b$ are dimensionless coefficients and represent the difference between the thermal pressure and the ram pressure, and gravity force term respectively. They checked the validity of the FEC in two different ways: 1) using the semi-analytic model of \cite{BURROWS1993}, 2) using 1D light-bulb simulations.  
They found that the analytic FEC is consistent with the semi-analytic explosion condition of \citet{BURROWS1993}. They also used this consistency to fit for the parameters $a$ and $b$ and find that $a = 0.06$ and $b= 0.38$. The light-bulb simulations also verified that the FEC is a useful diagnostic of explodability (\citet{Gogilashvili2022}, Figure 4).  In general, the simulations explode once they are above the explosion condition. In addition to providing an analytic explosion condition, the FEC also provides a nearness-to-explosion parameter.  For example, the simulations that start far away from the FEC never actually explode.  This suggests that one might be able to use the FEC as a nearness-to-explosion diagnostic and determine whether a potentially computationally-expensive simulation will actually explode or not. The success of FEC is promising in three ways.  One, the force explosion condition is analytic and helps to illuminate the underlying physics of explosions.  Two, the FEC is consistent with the critical neutrino luminoity condition and light-bulb simulations.  Three, this condition may be a useful explosion diagnostic for more realistic, three-dimensional radiation hydrodynamic core-collapse simulations.

While \citet{Gogilashvili2022} tested the FEC with one-dimensional light bulb simulations, in this manuscript, we show that the FEC is consistent with one-dimensional simulations that use actual  neutrino transport. To test this consistency, we also provide practical ways to calculate the FEC for CCSN simulations that use neutrino transport. 

The structure of this manuscript is as follows.  In section ~\ref{sec:simulaion} we introduce \texttt{GR1D}, a code that simulates CCSNe using  neutrino-transport  \citep{OCONNOR2010,OCONNOR2015}. In section~\ref{sec:pract_consideration}, we consider the FEC of \citet{Gogilashvili2022} in detail and introduce a practical FEC which can be used in more realistic, 1D radiation-hydrodynamic CCSN simulations. In section ~\ref{sec:verification} we test and verify the practical FEC using \texttt{GR1D}. We show that FEC accurately describes explodability for 1D radiation-hydrodynamics simulations. In section ~\ref{sec:discussion}, we discuss and summarize main conclusions and provide next steps for further using FEC.

\section{Spherically Symmetric Neutrino-transport Hydrodynamic Simulations (\texttt{GR1D})}\label{sec:simulaion}

We validate the FEC using the open-source, spherically symmetric, general-relativistic hydrodynamics, CCSN code \texttt{GR1D} \citep{OCONNOR2010,OCONNOR2015}. To facilitate time integration, \texttt{GR1D} slices space-time into space-like manifolds.  The manifolds are spherically symmetric such that the interval (and metric components) are  $ds^2=-\alpha(r,t)^2 dt^2+X(r,t)^2 dr^2+r^2d\Omega^2$, where $\alpha$ is the lapse and the radial metric component is $X^2=1/(1-2m(r)/r)$; $m(r)$ is the enclosed gravitational mass. \texttt{GR1D} solves the metric coefficients using Hamiltonian and momentum constraints, and it solves the hydrodynamic evolution equations via the Valencia formulation of relativistic hydrodynamics \citep{FONT2000}.  The set of evolution equations for matter is:
    \begin{equation}
        \partial_t \boldsymbol{U}+\frac{1}{r^2}\partial_r\big[\frac{\alpha r^2}{X}\boldsymbol{F}\big]=\boldsymbol{S} \, ,
    \end{equation}
where $\boldsymbol{U}$ is a set of conserved variables, $\boldsymbol{F}$ is their flux vector, and $\boldsymbol{S}$ is a vector of gravitational, geometric, and neutrino-matter interaction source terms.

The spatial discretization is finite-volume. The time integrator is a second order Runge-Kutta algorithm. To close the system of equations, \texttt{GR1D} includes several equations of state (EOS-s).  They include analytic EOS-s such as polytropic and $\Gamma$-law EOS-s. For CCSN simulations, the code can also handle the tabulated EOS, which includes relativistic and dense nuclear matter, for this work we use the SFHo EOS \cite{STEINER2013}. The code treats neutrino transport using the energy-dependent, truncated moment formalism of \citet{SHIBATA2011} and \citet{CARDALL2013}.  The moment equations assuming analytic closures for the higher moments. Neutrino-matter interactions are handled through \texttt{NuLib}.  \citet{OCONNOR2010} and \citet{OCONNOR2015} provide detailed descriptions and derivations of the equations and approaches in \texttt{GR1D}. 

This manuscript presents a systematic exploration of the explosion conditions for four progenitor models: $12M_\odot$, $15M_\odot$, $20M_\odot$, and $25M_\odot$ \citep{WOOSLEY2007,SUKHBOLD2014}. These models do not explode in self-consistent 1D simulations.  Therefore, to explore the explosion conditions, we modify the cross sections for electron-type neutrinos and anti-neutrinos
within the gain region.  Within the gain region (defined by the conditions of $\rho<3\times10^{10}\,\mathrm{g}\,\mathrm{cm}^{-3}$, $s>6\,k_\mathrm{B}/\mathrm{baryon}$, and $d\tau_\nu > 0$, where $d\tau_\nu$ is the change in the matter energy due to neutrino interactions), we enhance the energy absorption ($d\tau_\nu$) by a factor, $f_\mathrm{heat}$.
In this way, much of the neutrino transport is self-consistent, yet we also explore the explosion conditions in spherically symmetric simulations.    

\begin{figure*}
\centering
\captionsetup[subfigure]{labelformat=empty}
\begin{subfigure}[t]{\columnwidth}
\centering
\includegraphics[width=\linewidth]{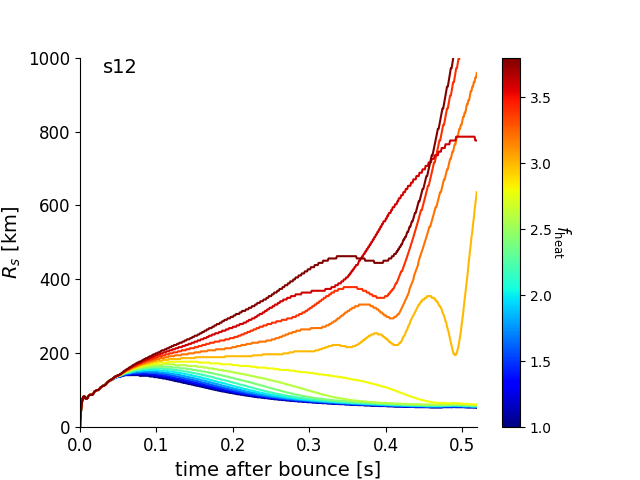}
\end{subfigure}
\begin{subfigure}[t]{\columnwidth}
\centering
\includegraphics[width=\linewidth]{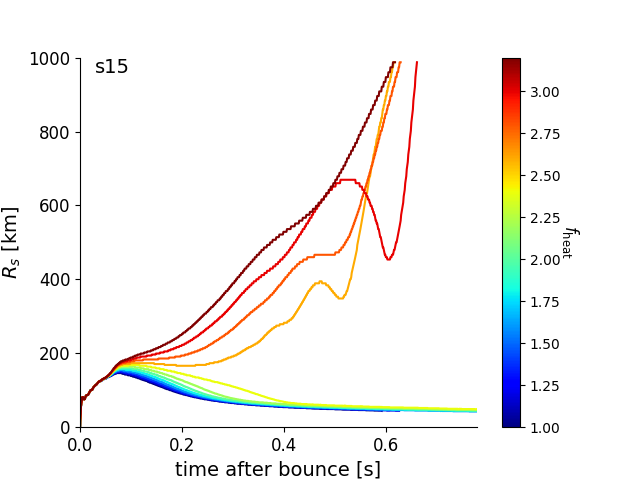}
\end{subfigure}

\begin{subfigure}[t]{\columnwidth}
\centering
\vspace{0pt}
\includegraphics[width=\linewidth]{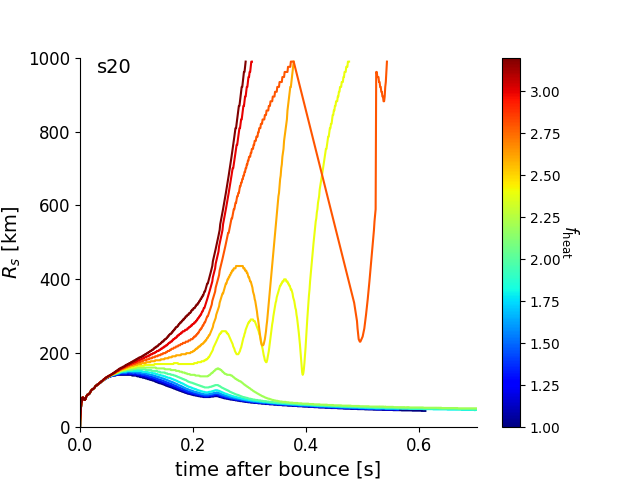}
\end{subfigure}
\begin{subfigure}[t]{\columnwidth}
\centering
\vspace{0pt}
\includegraphics[width=\linewidth]{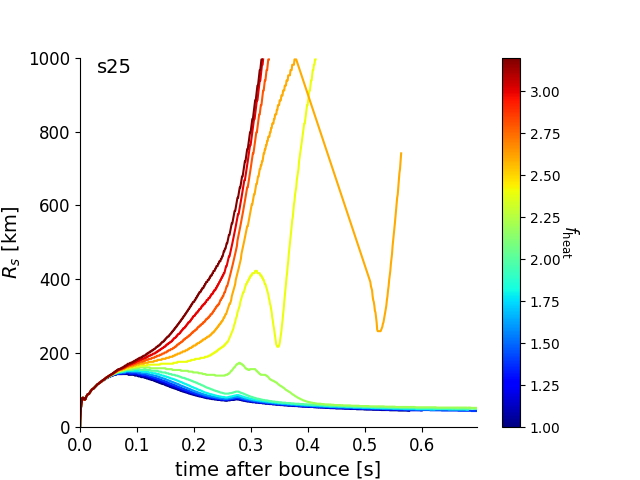}
\end{subfigure}
\begin{minipage}[t]{\textwidth}
\caption{Shock radius evolution for four  progenitors  \citep{SUKHBOLD2014}. For each progenitor, the colors delineate  the heating parameter. $12 M_\odot$ progenitor explodes for $ \rm f_{\rm heat} \geq 3.0$. $15 M_\odot$ progenitor explodes for $ \rm f_{\rm heat} \geq 2.6$ while $20 M_\odot$ and $25 M_\odot$ progenitors explode for $ \rm f_{\rm heat} \geq 2.4$.}
\label{fig:shock_vs_time} 
\end{minipage}
\end{figure*}

Figure~\ref{fig:shock_vs_time} shows the evolution of the shock radius for all four progenitors. For each progenitor, the colors represent different simulations with different heating factors. For the $12M_\odot$ progenitor, simulations with $\rm f_{\rm heat} < 3.0$ do not explode while simulations with $\rm f_{\rm heat} \geq 3.0$ explode. For the $15M_\odot$ progenitor, simulations with $ \rm f_{\rm heat} \geq 2.6$ explode. For the $20M_\odot$ and $25M_\odot$ progenitors, the simulations explode for $\rm f_{\rm heat} \geq 2.4$.

\section{Practical Considerations for Applying FEC to Spherical Radiation-Hydrodynamic Simulations}\label{sec:pract_consideration}

The analytic FEC requires some approximations for neutrino transport and the hydrodynamic structure.  All of these approximations are consistent with the light-bulb simulations but to test whether these approximations are also consistent with the neutrino transport, one must translate some of the approximate dimensional parameters into parameters that are practical for radiation-hydrodynamic simulations.  

In the analytic derivation, \citet{Gogilashvili2022} started from the fundamental equations of hydrodynamics and considered the balance of integral forces. They found that there are only two important dimensionless parameters determining explodability of the star. The first dimensionless parameter is the dimensionless neutrino heat deposited in the gain region, $\tilde{L}_\nu \tau_g = L_{\nu} \tau_g R_{\rm NS}/ ( G \dot{M} M_{\rm NS})$.  The second is the dimensionless neutrino opacity, $\tilde{\kappa} = \kappa \dot{M} / \sqrt{G M_{\rm NS} R_{\rm NS}}$, that parameterizes the neutrino optical depth in the accreted matter near the neutron-star surface. Here $L_\nu$ is neutrino luminosity, $\tau_g$ is the neutrino optical depth within the gain region, $\dot{M}$ is the mass accretion rate, $R_{\rm NS}$ and $M_{\rm NS}$ are the size and mass of the proto-neutron star (PNS). Within an explicit set of self-consistent approximations, \citet{Gogilashvili2022} derived that the FEC is $\tilde{L}_\nu \tau_g -0.06\tilde{\kappa} = 0.38$. They validated and tested the FEC in two ways. In the first, they validated the FEC with the steady-state solutions of \citet{BURROWS1993}. In the second, they tested the FEC using 1D light-bulb simulations. \citet{Gogilashvili2022} showed that the FEC reproduces the explosion conditions, while simulations that do not satisfy the FEC fail to explode. Moreover, they showed that the FEC is an accurate explosion diagnostic for 1D light-bulb simulations and proposed that it might be a useful diagnostic for more-realistic, one-dimensional and multi-dimensional simulations.

\begin{figure}
    \centering
    \includegraphics[width=\columnwidth]{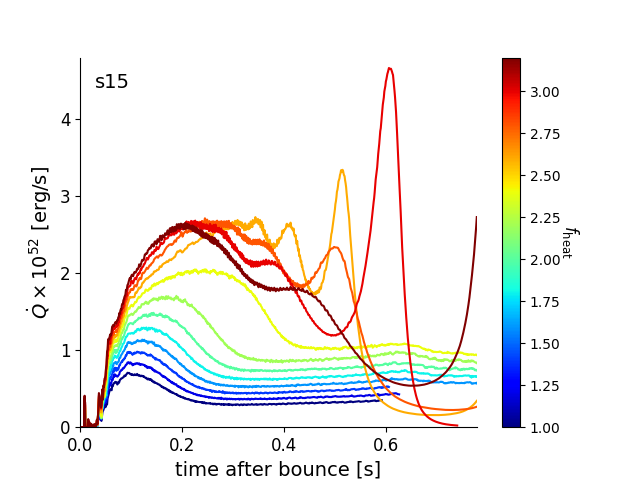}
    \caption{Evolution of total net neutrino heating deposited in the gain region for the $15 M_\odot$ progenitor  \citep{SUKHBOLD2014}. The colors indicate the heating parameter.    In the analytic FEC, an important dimensionless parameter is $\tilde{L}_{\nu} \tau = L_{\nu} \tau R_{\rm NS}/ (G \dot{M} M_{\rm NS})$.  To make the FEC a practical explosion condition for actual neutrino transport, we propose to replace $L_{\nu} \tau$ with $\dot{Q}$, the net neutrino power deposited in the gain region.}
    \label{fig:Heat_comparison}
\end{figure}

In this section, we propose slight, practical modifications to the FEC that can be applied to spherically symmetric, radiation-hydrodynamic simulations. One of the important dimensionful parameters in the FEC is $L_\nu \tau_g$. A more practical parameter is to replace $L_\nu \tau_g$ with the net neutrino heating deposited in the gain region, $\dot{Q}$.  This is after all the intent and definition of $L_{\nu} \tau_g$, and one can easily calculate the net neutrino heating within the gain region.  Figure~\ref{fig:Heat_comparison} shows the evolution of $\dot{Q}$ for the $15 M_\odot$ progenitor \citep{SUKHBOLD2014}. The colors indicate different simulations for the different heating parameters. 

\begin{figure}
\includegraphics[width=\columnwidth]{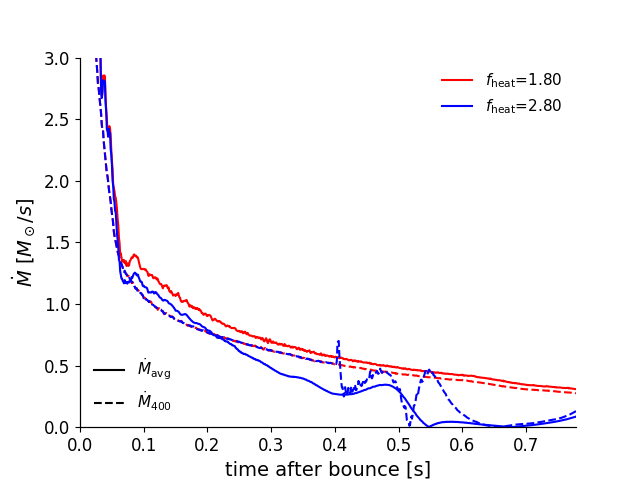}
\caption{Mass accretion rate vs.~time, $\dot{M}$, for the $15~M_\odot$ progenitor \citep{SUKHBOLD2014}. The red lines show $ \rm f_{\rm heat}=1.80$, which does not explode and the blue lines show $ \rm f_{\rm heat}=2.80$, which does explode. The dashed lines show $\dot{M}$ at $400 km$, while the solid lines show $\dot{M}$ averaged over the gain region. For the non-exploding model, there is little difference between $\dot{M}(400 \rm{km})$ and $\dot{M}$ within the gain region.  However, for the exploding model, $\dot{M}$ within the gain region and above the shock are not the same.  Since $\dot{M}$ within the gain region determines the residency and heating within the gain region, $\dot{M}$ within the gain region is the most accurate estimate when calculating the FEC. }
\label{fig_mdot_comparison}
\end{figure}

\begin{figure}
    \centering
    \includegraphics[width=\columnwidth]{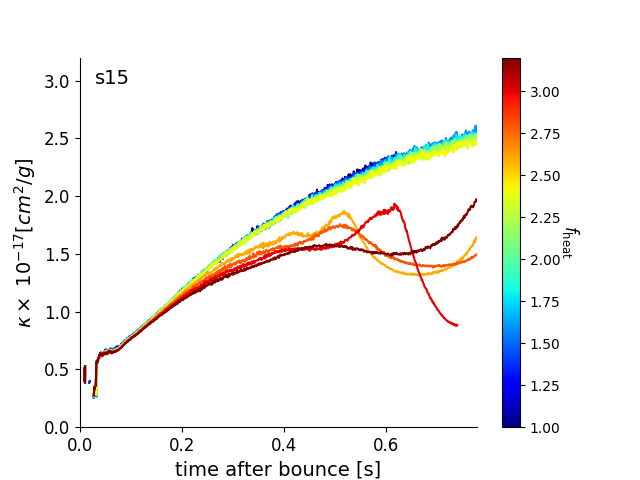}
    \caption{The luminosity-weighted neutrino opacity at the gain radius as a function of time. In the Force Explosion Condition, neutrino opacity is one of the important parameters. For one-dimensional simulations that include neutrino transport, we propose to use this energy density-averaged neutrino opacity measured at the gain radius. 
    }
    \label{fig:kappa}
\end{figure}

Another important dimensionful parameter in the FEC is the mass accretion rate, $\dot{M}$. Figure~\ref{fig_mdot_comparison} illustrates a slight ambiguity in choosing where to measure $\dot{M}$.  Within the steady-state assumption, $\dot{M}$ is constant in both time and radius.  In actual simulations, this is a good approximation most of the time, especially while the shock is stalled.  However, there is a phase in which the assumption that it is constant in radius breaks down.  Just as the simulation is about to explode it enters a phase in which the shock expands slower than it will when it begins to explode vigorously.  Even though the expansion is not as vigorous, it still violates the assumption that $\dot{M}$ is independent of radius.  Figure~\ref{fig_mdot_comparison} shows the evolution of $\dot{M}$ for a $15 M_\odot$ progenitor.
The red lines correspond to $f_{\rm heat}=1.80$ simulation, which does not explode, and the blue lines correspond to $f_{\rm heat}=2.80$ simulation, which does explode. For each simulation, we compare the mass accretion rate calculated at $400$\,km ($\dot{M}_{400}$; dashed lines) and the mass accretion rate averaged within the gain region ($\dot{M}_{\rm avg}$; solid lines).  For the non-exploding model, $\dot{M}_{400}$ and the gain region-averaged $\dot{M}_{\rm avg}$ differ only by a few percent. However, for the exploding model, $\dot{M}_{400}$ begins to deviate from $\dot{M}_{\rm avg}$ by large factors.  Even while the shock is moving outward slowly, $\dot{M}_{\rm avg}$ is less than $\dot{M}_{400}$.  Since it is the mass accretion rate within the gain region that determines the dwell time and hence heating within the gain region, we propose using $\dot{M}_{\rm avg}$ as the appropriate measure of $\dot{M}$ for the FEC. 

The neutrino opacity also plays an important role in the critical condition.  In the analytic FEC, \citet{Gogilashvili2022} assumed a single temperature for the neutrinos being captured within the gain region.  Hence, \citet{Gogilashvili2022} assumed a single opacity for the gain region.  In reality, the effective temperature of the neutrino distribution changes with radius.  Therefore, we must choose an appropriate temperature and opacity for the practical FEC. For one-dimensional simulations that include neutrino transport, we propose using a weighted average of the electron-neutrino and the electron-antineutrino opacities  measured at the gain radius. First, we calculated the neutrino-energy-density weighted opacities for each neutrino species; they are $\kappa_{\nu_e}$ and $\kappa_{\overline{\nu}_e}$, respectively.  We then calculate the luminosity weighted opacities as $\kappa=(L_{\nu_e}\kappa_{\nu_e}+L_{\overline{\nu}_e}\kappa_{\overline{\nu}_e})/(L_{\nu_e}+L_{\overline{\nu}_e})$.  Figure~\ref{fig:kappa} shows the time evolution of this luminosity-weighted opacity of the neutrinos and anti-neutrinos measured at the gain radius.  Again, the colors represent different heating factors. The neutrino opacity at the gain radius is an accurate estimate when calculating the FEC.

Therefore, a practical FEC that can be applied to one-dimensional radiation-hydrodynamics simulations is $\tilde{\dot{Q}}-0.06\tilde{\kappa} = 0.38$, where $\tilde{\dot{Q}}=\dot{Q}R_{\rm NS}/ ( G \dot{M} M_{\rm NS})$ is the dimensionless neutrino heating deposited in the gain region and $\tilde{\kappa} = \kappa \dot{M} / \sqrt{G M_{\rm NS} R_{\rm NS}}$ is the dimensionless, luminosity-averaged neutrino opacity at the gain radius.

\begin{figure*}
\centering
\captionsetup[subfigure]{labelformat=empty}
\begin{subfigure}[t]{\columnwidth}
\centering
\includegraphics[width=\linewidth]{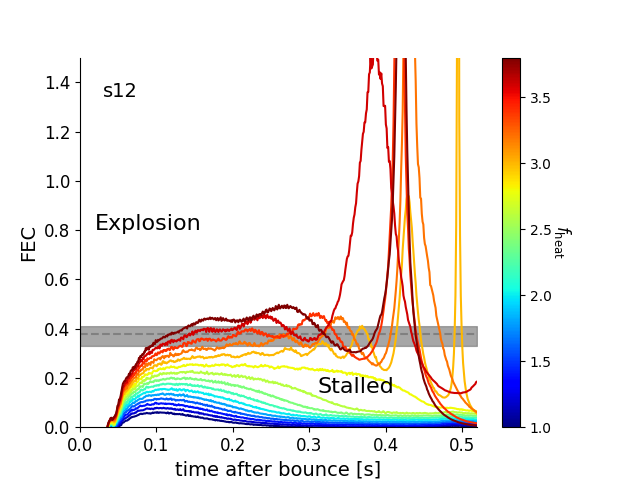}
\end{subfigure}
\begin{subfigure}[t]{\columnwidth}
\centering
\includegraphics[width=\linewidth]{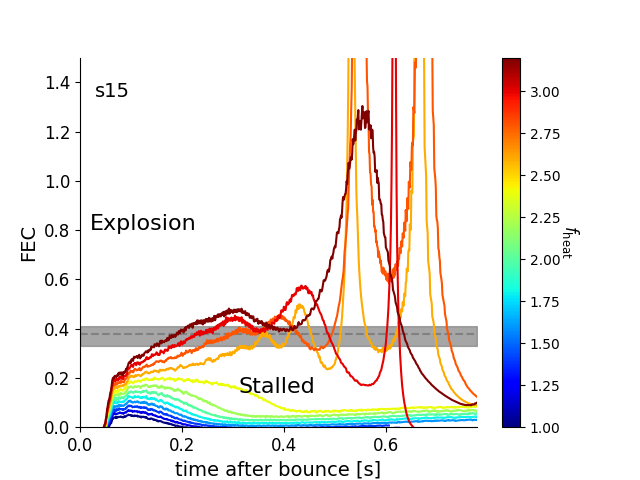}
\end{subfigure}

\begin{subfigure}[t]{\columnwidth}
\centering
\vspace{0pt}
\includegraphics[width=\linewidth]{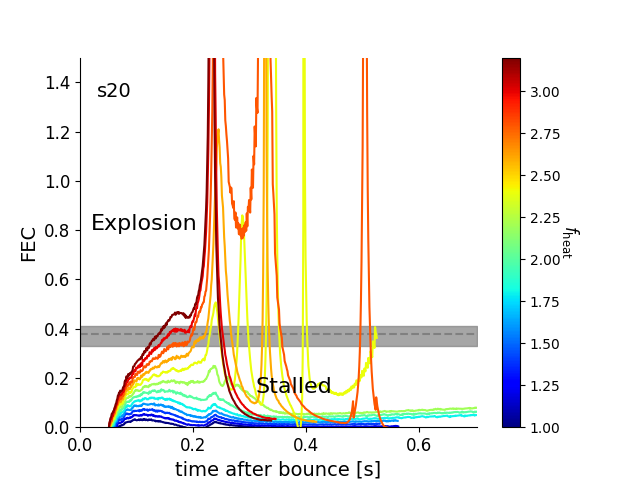}
\end{subfigure}
\begin{subfigure}[t]{\columnwidth}
\centering
\vspace{0pt}
\includegraphics[width=\linewidth]{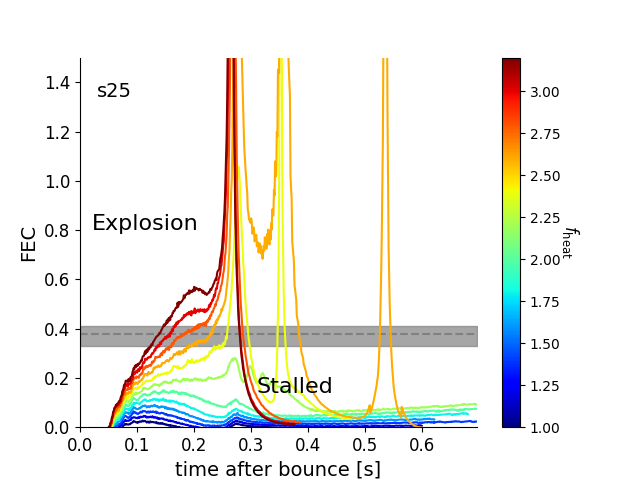}
\end{subfigure}
\begin{minipage}[t]{\textwidth}
\caption{Evolution of FEC for $12 M_\odot$, $15 M_\odot$, $20 M_\odot$, and $25 M_\odot$ progenitors of \citet{SUKHBOLD2014}. For each progenitor, the colors represent the $\rm f_{\rm heat}$. The dashed line shows the best fit critical condition from the steady-state solutions and the gray band shows the variation in the critical condition based upon the steady-state solutions \citep{Gogilashvili2022}; For each progenitor, the simulations that cross the gray band explode, while the simulations that do not cross the band - fail to explode.   The gray band represents a prediction from \citet{Gogilashvili2022} that was based upon some strong assumptions about neutrino transport.  The fact that the FEC is consistent with one-dimensional radiation-hydrodynamic simulations is encouraging for the FEC as a theory.}
\label{fig:FEC_vs_time}
\end{minipage}

\end{figure*}

\section{Verifying The Practical Explosion Condition With \texttt{GR1D} Simulations}\label{sec:verification}
In this section, we use \texttt{GR1D} simulations to test and validate the FEC. We simulate four different progenitors: $12M_\odot$, $15M_\odot$, $20M_\odot$, and $25M_\odot$ \citep{SUKHBOLD2014}. For each progenitor, we vary the heating factor, $f_{\rm heat}$.

Next, using the practical FEC dimensionless parameters (section \ref{sec:pract_consideration}), we calculate the time evolution of the FEC; see Figure~\ref{fig:FEC_vs_time}. Again, the colors indicate $f_{\rm heat}$. The dashed line shows the best fit for critical condition from the steady-state solutions and the gray band shows the variance in the critical condition \citep{Gogilashvili2022}.  For each progenitor, the simulations that explode cross the gray band while the simulations that do not explode never cross the gray band. For a detailed analysis of this result, consider the results for the  $12\,M_\odot$ progenitor (left top panel). For this progenitor, $f_{\rm heat}\geq 3.0$ (orange and red lines) eventually leads to explosion. For the $f_{\rm heat}=3.0$ simulation, the shock radius slowly increases (with slight oscillations) starting at 100\,ms after bounce until $\sim$500\,ms after bounce when the shock expansion rapidly accelerates (see Figure \ref{fig:shock_vs_time}). Figure~\ref{fig:FEC_vs_time} shows that the FEC evolution coincides with the shock evolution and is a good indicator of explosion.  In particular, the FEC reaches $\sim$0.3 at 100\,ms and slowly increases through the grey band over the following 100s of ms.  The simulations with $f_{\rm heat}\leq2.8$, which do not explode, are comfortably below the grey band for the entire simulation. The simulations involving the other progenitors ($15M_\odot$, $20M_\odot$, and $25M_\odot$) show the same qualitative results in that the FEC is a good indicator of explosion.   Similarly to the $12M_\odot$ progenitor, the $15M_\odot$ progenitor explodes later around $\sim 500\,\mathrm{ms}$ after bounce. However, for the $15M_\odot$ progenitor, the minimum $f_{\rm heat}$ to explode is 2.6.   The $20M_\odot$ and $25M_\odot$ progenitors explode earlier ($\sim 300\,\mathrm{ms}$ after bounce). For both $20M_\odot$ and $25M_\odot$ progenitors the explosion is realized for $f_{\rm heat}\geq 2.4$.  These results show that the FEC is consistent with these one-dimensional radiation-hydrodynamic simulations. Therefore, we conclude that the FEC is indeed an accurate explosion diagnostic for spherical simulations. In addition to successfully reproducing the explosion conditions, one may use the FEC to calculate the distance from the explosion. 

As an aside, as is often the case \citep{MURPHY2008, MABANTA2019,Gogilashvili2022}, these 1D simulations show oscillations in the simulations that are near explosion.  As yet, no one has provided a clear and definitive explanation for this oscillations.  Interestingly, The timescale for the oscillations is on the order of the advection timescale (10s of ms) and not the sound crossing timescale.  Such an oscillation is reminiscent of the advective-acoustic cycle (private communication with 
T. Foglizzo, \citet{Foglizzo2007}).  To date, most analyses of the advective-acoustic cycle are focused on nonradial oscillations.  These oscillations are clearly spherical in nature, and we are left wondering if they are in some way connected to the advective-acoustic cycle.   Nevertheless, these oscillations do not alter the qualitative and broadly quantitative conclusion that the FEC is consistent with the explosion conditions of the 1D radiation hydrodynamic simulations.

\section{Discussion and Conclusions}\label{sec:discussion}
Being able to predict which stars explode, leaving behind neutron stars and which collapse to black holes is important in understanding the final fates of massive stars.  These predictions will require both numerical and analytic theory.  The analytic force explosion condition (FEC) \citep{Gogilashvili2022} shows promise in that it is consistent with the explosion conditions of spherically symmetric light-bulb simulations.  To test whether the FEC is also consistent with spherically symmetric neutrino transport, we propose practical modifications to the dimensionless parameters of the FEC.  Using \texttt{GR1D} simulations, we find that the FEC is indeed consistent with the explosion conditions of  spherically symmetric, radiation-hydrodynamic, CCSN simulations. Moreover, one may use the FEC to determine a distance to explosion. 
These results further suggest that the Force Explosion Condition (FEC) may accurately describe the explosion conditions of two- and three- dimensional CCSN simulations.  This will require including an analytic mean-field model for neutrino-driven convection \citep{MABANTA2018}. Of course the final test will require comparing this FEC plus mean-field convection with 2D and 3D radiation-hydrodynamic simulations.

\section*{Acknowledgements}
MG and JWM are supported by the Los Alamos National Laboratory (LANL) through its Center for Space and Earth Science (CSES). CSES is funded by LANL’s Laboratory Directed Research and Development (LDRD) program under project number 20210528CR. EO is supported by the Swedish Research Council (Project No. 2020-00452).  Computations were enabled by resources provided by the Swedish National Infrastructure for Computing (SNIC), partially funded by the Swedish Research Council through grant agreement no. 2018-05973.

\section*{Data Availability}
The data underlying this article will be shared on reasonable request to the corresponding author.

\medskip
\label{lastpage}
\bibliography{References}

\end{document}